\documentclass[%
superscriptaddress,
unsortedaddress,
 amsmath,amssymb,
 aps,
twocolumn
]{revtex4-1}

\usepackage{graphicx}
\usepackage{dcolumn}
\usepackage{bm}
\usepackage{xcolor,soul}

\usepackage{hyperref}

\DeclareRobustCommand{\vect}[1]{
  \ifcat#1\relax
    \boldsymbol{#1}
  \else
    \mathbf{#1}
  \fi}

  \newcommand{\cbr}[1]{\left(#1\right)}
  \newcommand{\sbr}[1]{\left[#1\right]}

\begin{document}

\title{Self-diffusion in a strongly coupled non-neutral plasma}

\author{Marco Baldovin}
 \email{marco.baldovin@cnr.it}
\affiliation{Institute for Complex Systems, CNR, 00185, Rome, Italy}
\affiliation{Universit\'{e}  Paris-Saclay,  CNRS,  LPTMS,  91405,  Orsay,  France}

 \author{Gr\'egoire Vallet}
 \affiliation{Aix-Marseille University, CNRS, PIIM, Marseille, France}

 \author{Ga\"{e}tan Hagel}
 \affiliation{Aix-Marseille University, CNRS, PIIM, Marseille, France}

 \author{Emmanuel Trizac}%
\affiliation{Universit\'{e}  Paris-Saclay,  CNRS,  LPTMS,  91405,  Orsay,  France}%
\affiliation{ENS de Lyon, 69342 Lyon, France}
 
 \author{Caroline Champenois}
 \affiliation{Aix-Marseille University, CNRS, PIIM, Marseille, France}

\date{\today}

\begin{abstract}
We propose 
a joint experimental and theoretical approach
to measure the self-diffusion in a laser-cooled trapped ion cloud where part of the ions are shelved in a long-lived dark state. 
The role of the self-diffusion coefficient in the spatial organisation of the ions is deciphered, following from the 
good agreement between the experimental observations and the theoretical predictions. 
This comparison furthermore allows to deduce the temperature of the sample. Protocols to measure the self-diffusion coefficient are discussed, in regard with the control that can be reached on the relevant time scales through the dressing of the atomic levels by laser fields.
\end{abstract}

\maketitle


\section{Introduction}

Laser-cooled clouds of atomic ions stored in a radio-frequency trap are practical realisations of a finite-size One Component Plasma (OCP) in the strongly coupled regime. 
The OCP is a reference model in the study of strongly coupled Coulomb systems \cite{ichimaru82}. By tuning the density $n_i$ and the temperature $T$ of the sample, different regimes can be explored from gas to liquid and crystals. 
Standard kinetic theories \cite{stanton16}  fail to describe transport plasma properties under conditions of strong Coulomb coupling because they neglect effects of spatial and temporal correlations induced by nonbinary collisions \cite{strickler16}. This fundamental problem needs to be solved to accurately model the transport properties, and equations of state of dense laboratory and astrophysical plasmas.

Even if measurements are important to benchmark potential models and test plasma theories out of the conventional plasma regimes \cite{baalrud13,scheiner19,baalrud19}, few experiments can give access to relevant diffusivity parameters like the ion self-diffusion constant. The main contributions so far have come from experiments based on ultra-cold neutral plasmas created by photo-ionisation of an ultra-cold atomic cloud \cite{strickler16}. Despite the short lifetime of these neutral plasmas, the experimental data allow for studying the effects of strong coupling on collisional processes, which is of interest for dense laboratory and astrophysical plasmas \cite{strickler16,langin19, gorman22}. Here, we propose to use another model system for a strongly correlated plasma, which benefits from an infinitely long trapping lifetime. The presence of the external confining potential induces a finite size for the system, while Coulomb repulsion forces tend to maximize the mutual distance between ions. The interplay between these two effects typically results in a shell structure~\cite{Wrighton10, Chepelianskii2011}, which can be fairly modeled by a regular lattice geometry. This will be the starting point for the building of our model.

The next section introduces the experimental set-up and the characteristics of the system used. In section~\ref{sec:model}, we propose a model for the diffusion process, to identify the parameters that are accessible to experiments. Section~\ref{sec:compare_exp} compares the predictions of the model for the stationary regime to the results of the experiments. Section~\ref{sec:strategies} proposes experimental strategies to measure the diffusivity of the ions, while conclusions are drawn in section \ref{sec:concl}.

\section{Experimental setup}
\label{sec:exp}

 In the experiment considered as a support for studying diffusion properties,  clouds of few hundreds to thousands Ca$^+$  ions are   stored in a linear radio-frequency (rf) quadrupole  trap  where the role of the neutralising particles is played by the confining potential \cite{champenois09}. The technical details concerning the set-up can be found in \cite{champenois13,kamsap15t} and we recall here the useful facts. In the pseudo-potential approximation \cite{dehmelt67,gerlich92} which is relevant in the context of these experiments, the effective trapping potential can be described by
\begin{equation}
    V_{\text{trap}}(x,y,z)=\frac{1}{2}m\omega_r^2 (x^2+y^2)+\frac{1}{2}m\omega_z^2 z^2
\end{equation}
with $m$ the mass of a single ion, and $(x,y,z)$ its Cartesian coordinates. The potential depth is of several eV and its cylindrical symmetry is related to the quadrupole geometry which is built on four electrodes along the direction $Oz$. By means of Doppler laser-cooling,  temperatures $T$ of the order of 1 to 100~mK can be reached \cite{drewsen98,hornekaer02}. The density $n_i$ of the cloud is controlled by the strength of the rf trapping field, and scales with $\omega_r^2$. Through the tuning of the density and of the temperature, the plasma parameter 
\begin{equation}
   \mathcal{G}_p=\frac{q^2}{4\pi \epsilon_0 a k_B T}
\end{equation}
can be tuned over several orders of magnitude, where $a$ is the Wigner-Seitz cell radius defined as $(3/4\pi n_i)^{(1/3)}$, when $n_i$ has reached the cold limit for the density \cite{hornekaer02}, $q$ the charge of the Ca$^+$  ion and  $k_B$  the Boltzmann constant. It is possible to assign a gas, liquid and crystal state to such a sample, using the two body correlation function \cite{hansen73, pollock73}.
With the control on Doppler laser-cooling and on the steepness of the trapping potential, the plasma parameter of a trapped-ion based finite OCP can span from gas ($\mathcal{G}_p$ lower than 0.1) to liquid ($\mathcal{G}_p$ of the order of 1 to 100) and crystal phases  ($\mathcal{G}_p$  larger than 200). For temperatures lower than 1~K, the thermal kinetic energy is small compared to the trapping and Coulomb repulsion potential energies and  ions  arrange in a stationary structure that minimise the total potential energy, to form what is called a Coulomb crystal, of an ellipsoidal shape, characterised by a radius $R$ and length $L$, with an aspect ratio $R/L$ controlled by the trapping potential aspect ratio $\omega_r^2/\omega_z^2$ \cite{hornekaer02}. 
 \begin{figure}
    \centering
   \includegraphics[width=\linewidth]{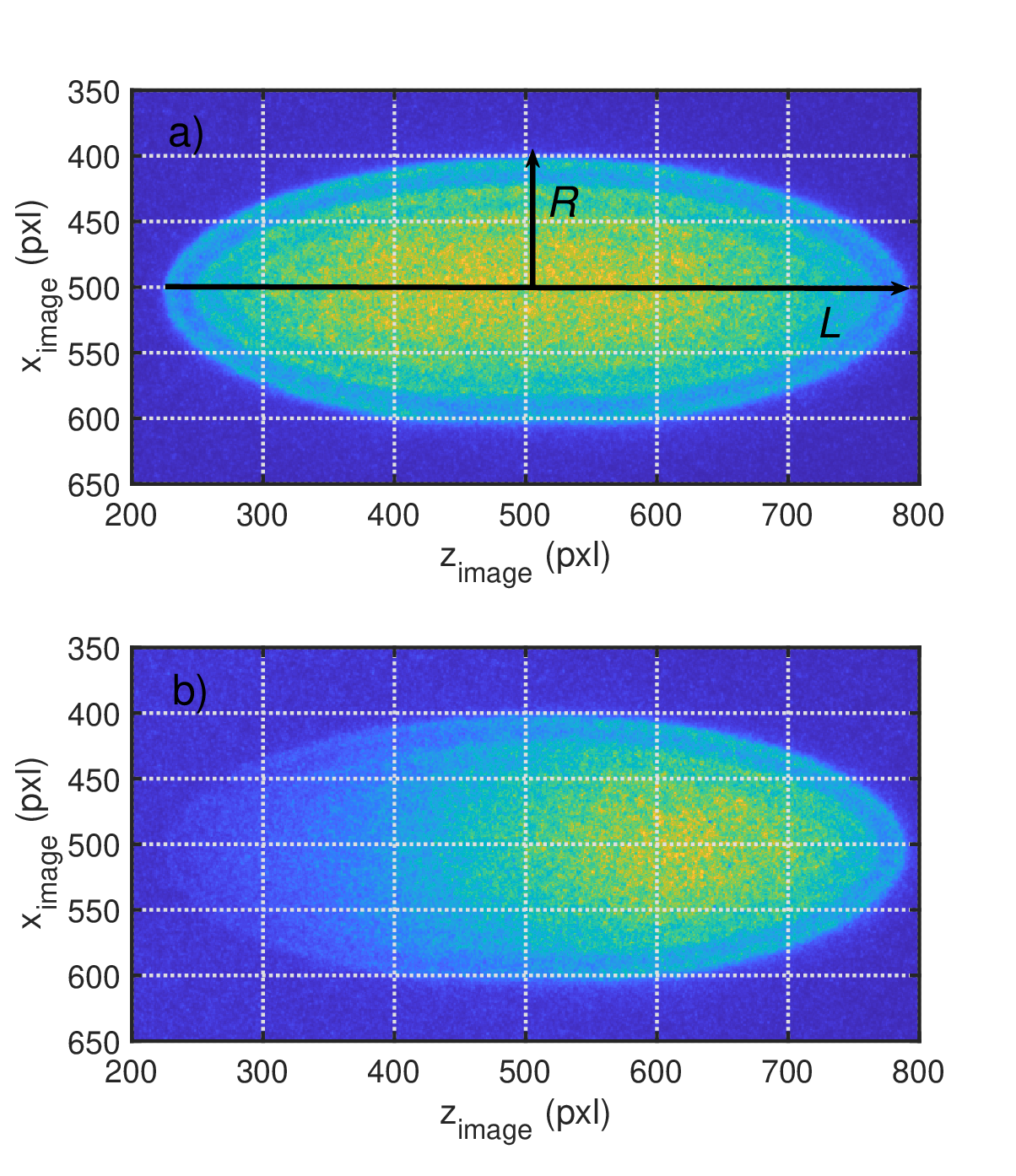}
    \caption{Panel (a): picture of the laser induced fluorescence of a trapped ion cloud made of 1240~$\pm 50$ Ca$^+$ ions.  Given an optical magnification of  $12.6\pm 0.1$ and a pixel size of 13~$\mu$m, the scale of the picture in pxl is very close to the scale of the cloud in $\mu$m. The cloud dimensions  are : $R=107 (\pm 2) \mu$m for the largest radius and $L= 576(\pm 14)\mu $m for the total length. Panel (b) : same cloud with an extra laser admitted, propagating toward $z>0$ in real space and on the picture, this laser is shelving part of the ions in a metastable dark state.}
    \label{fig:picture}
\end{figure}

One can show \cite{prasad79, dubin99,hornekaer02, champenois09} that starting from a gaseous (``high temperature'') region and cooling to the liquid phase, the ion density is uniform in the cloud, except for an outside layer, of a width of the order of  few $\mu$m, small compared to the sizes of the cloud which are  of the order of several hundreds of $\mu$m (see Fig.~\ref{fig:picture}). When cooled further to the crystal phase, the outer shape of the cloud does not change and the mean density remains equal as in the liquid phase. It is of the order of $10^8$~cm$^{-3}$ for the experiments mentioned in the following. An example of this structure formed by 1240~$\pm 50$ ions is shown on Fig.~\ref{fig:picture}a, obtained by the image of their laser induced fluorescence on an intensified CCD camera. The fluorescence is driven by the laser excitation used for Doppler cooling. In the case of such 3D-trapped sample, a single laser beam propagating along the $Oz$ direction is sufficient for Doppler laser cooling. Here it propagates  toward the positive $z$. The pixel signal is proportional to the number of emitted photons integrated along the line of sight, which is one of the direction perpendicular to the trap axis $Oz$.

Such systems can be considered as a finite size realisation of a one-component plasma and we can get an estimate of the self-diffusion coefficient $D$ from the work of Daligault, based on molecular dynamics simulations \cite{daligault12} or computation by practical model \cite{daligault12b}. For the typical range of parameters covered by the experiments detailed in the following, these results lead to a value for $D$ of the order of 1 to $10\times 10^{-6}$~m$^2/s$.

Introducing extra lasers, the internal structure of   Ca$^+$ allows to shelve the electronic state in a metastable state \cite{nagourney86,knoop04}, or to trap it in a  dark state involving a coherent superposition of  two \cite{janik85,lisowski05} or three  \cite{champenois06,collombon19b} stable and metastable states. In these three situations, the ions cannot be excited by the cooling laser and thus do not scatter photons. These dark states have a lifetime in the millisecond to second range, which is far longer than the 6.9~ns lifetime of the excited state involved in the Doppler laser cooling. Because of several experimental imperfection like Doppler effect or collisions, the shelving or coherent trapping process does not involve all the ions at the same time and  some ions are still enrolled in the laser cooling process, inducing a cycling of absorption and spontaneous emission. The net recoil induced by each cycle is responsible for a radiation pressure \cite{metcalfbook} that is applied only on these bright ions. This state selective force is responsible for the spatial segregation between the bright and dark ions that is visible on figure~\ref{fig:picture}b. In spite of this,  thermalisation between the dark ions and the laser cooled ones allows to keep the cloud in a steady state, where segregation by the state-selective radiation pressure offers a unique tool to measure diffusion properties within a strongly correlated non-neutral plasma.

\section{Analytical model}\label{sec:model}
In this section we propose a simplified model for the considered system, assuming that the dynamics can be mapped,  at a microscopical level, to an exclusion process on a lattice. Such a regular geometry is inspired by the ordered structure at short scale of OCP~\cite{Wrighton10}. The dynamics does not allow the simultaneous presence of more than one ion in a ``site'' of the lattice, hence the restriction to an exclusion dynamics, in the spirit of the celebrated Asymmetric Exclusion Process~\cite{derrida1998exactly}.  At first we will neglect the internal state long lived dynamics, i.e. we will assume that the ions are not allowed to pass from the bright to the dark state and vice-versa. We will say that the states of the ions are ``frozen''. This simplifying hypothesis accounts for the fact that the typical life-times of the shelved states are much longer than the characteristic times for the spatial displacement of the ions. While the main features of the model will be already present in the one-dimensional setting, an accurate functional form to be compared with experimental density profiles will be provided by considering a more realistic  ellipsoidal geometry. In subsection~\ref{sec:shelving} the consequences of a non-frozen regime will be discussed, as to provide strategies for future experiments in this regime also.

\subsection{Limit of infinite lifetime of the dark state}\label{sec:frozen}

\subsubsection{Derivation of the model}
We consider a one-dimensional lattice, extending along the $z$ axis, composed by $N$ cells with size $a$. Each site of the lattice is occupied by an ion, and each ion can be found either in a bright or  in a dark state. For the moment, let us assume that transitions to and from the metastable state do not occur on the time-scale of the observed dynamics (frozen states).  The only allowed evolution is the swapping of neighbour particles  along the $z$ direction. We will denote by $\gamma_u$ the rate at which a bright particle placed on site $n$ exchanges its position with a dark particle at site $n+1$, by $\gamma_d$ the exchange rate toward site $n-1$. The two rates are determined by thermal fluctuations and by the effect of radiation pressure, as it will be discussed in the following. In particular, we expect $\gamma_u$ and $\gamma_d$ to be identical in the absence of external forces; the cooling laser along the $z$ axis, acting only on the bright ions, leads instead to unbalance.

Let us denote by $p_n(t)$  the probability that, at time $t$, the site $n$  is occupied by a bright particle. We can write down an evolution equation for the $\{p_n\}$, $1<n<N$ by recalling the swapping rules introduced before:
\begin{equation}
\label{eq:evw}
\begin{aligned}
 \frac{d p_n}{dt}=&-\gamma_u p_n (1-p_{n+1})-\gamma_d p_n (1-p_{n-1})\\
 &+\gamma_u p_{n-1} (1-p_{n})+\gamma_d p_{n+1} (1-p_{n})\,.
\end{aligned}     
\end{equation}
The first two terms on the right hand side are loss terms, as they account for the cases in which a bright particle initially present in the $n$th site leaves it and goes to a neighbour site, occupied by a dark ion; the two gain terms stand for the opposite transitions. The evolution equations are completed by the boundary conditions
\begin{equation}
\begin{aligned}
 \frac{d p_1}{dt}&=-\gamma_u p_1 (1-p_{2})+\gamma_d p_{2} (1-p_{1})\,\\
 \frac{d p_N}{dt}&=-\gamma_d p_N (1-p_{N-1})+\gamma_u p_{N-1} (1-p_{N})\,.
 \end{aligned}
\end{equation}

This scenario relies on the approximation that $p_n$ and $p_{n+1}$ are independent probabilities. A more accurate description of the system would involve the conditional probabilities of finding bright particles in the neighbor sites $n-1$ and $n+1$, given the occupation in the  site $n$. The equations ruling the time evolution of such two-site probabilities would requires three-site terms, leading to a hierarchy of coupled equations hardly addressable. The factorization hypothesis, which is reminiscent of Boltzmann's \textit{Stosszahlansatz} (``molecular chaos'' hypothesis), allows to close the equations at the first level of the hierarchy (see, e.g., Chapter 3.3 of Ref.~\cite{huang87}). The quality of the approximation is checked \textit{a posteriori}, by comparing the predictions of the model with the experimental results.

\subsubsection{Physical interpretation}

It is useful to switch to a continuous description, as it is typically done when considering the hydrodynamic behaviour of a gas of particles. Passing to such a coarse-grained model yields the twofold benefit of (i) getting insight about the physical interpretation of the terms ruling the evolution and (ii) finding explicitly -- at least in some cases -- the stationary state. The price to pay is a lower accuracy on the short length-scales. To this end, we define a density of bright particles as
$$
\int_0^z d z'\rho(z')=\sum_{n=1}^{\lfloor z/a \rfloor}p_n\,,
$$
where $\lfloor x \rfloor$ denotes the largest integer smaller than $x$. We are interested in the limit $N \gg 1$, with the total length of the lattice, $L=Na$, finite. We can then derive an evolution equation for the density from Eq.~\eqref{eq:evw}, by substituting
$$
\begin{aligned}
p_{n}&\to a \rho(z)\\
p_{n \pm 1} &\to a \rho(z \pm a)\simeq a \rho(z) \pm a^2 \partial_z\rho(z)+ \frac{a^3}{2} \partial^2_z\rho(z) + ...
\end{aligned}
$$

One gets 
\begin{equation}
\label{eq:evrho}
\partial_t \rho = - V\partial_z\rho +2aV\rho \partial_z \rho + D\partial^2_z\rho\,,
\end{equation} 
where
\begin{equation}
\label{eq:coeff}
V= a(\gamma_u-\gamma_d)\quad \quad D=\frac{a^2(\gamma_u+\gamma_d)}{2}\,.
\end{equation}
The evolution equation~\eqref{eq:evrho} is a Burgers' equation with viscosity~\cite{Ablowitz2011-ww}, often encountered when taking the hydrodynamic limit of asymmetric exclusion processes~\cite{Mallick2015, Lazarescu2015}. The first term on the r.h.s. is a systematic drift due to the presence of the external forcing by the radiation pressure; the second one, nonlinear in the density, accounts for the exclusion processes that favor the occupation of a site by a single particle; the last term, proportional to the second derivative of the density, accounts for diffusion. Equation~\eqref{eq:evrho} can be written in the form of a conservation law
\begin{equation}
    \partial_t \rho=-\partial_z J(z)\,,
\end{equation}
where
\begin{equation}
   J(z)=V\rho(z) -aV\rho^2(z)-D\partial_z \rho(z)
   \label{eq:current}
\end{equation}
plays the role of a density current.

\begin{figure}
 \centering\includegraphics[width=\linewidth]{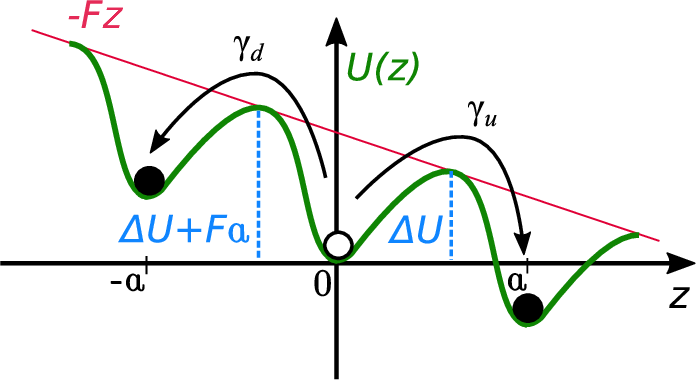}
 \caption{Sketch of the potential felt by a bright particle moving along the lattice. The tilt of the periodic potential (green curve) is induced by a force $F$ pointing toward $z>0$ (red line).  \label{fig:potential}}
\end{figure}

A physical interpretation of the above scenario can be obtained by considering an effective potential $U(z)$ felt by a bright particle moving in the lattice, as sketched in Fig.~\ref{fig:potential}.
This potential accounts for the repulsion between adjacent particles and for the effect of the external force $F$ pointing along $z>0$ which tilts the potential profile.  Let us call $\Delta U$ the potential barrier that a bright particle placed in $z$ needs to overcome in order to swap with a neighbour dark particle placed in $z+a$. Because of the tilt, the potential barrier associated with a swapping with a dark particle in $z-a$ is thus $\Delta U + Fa$. An Arrhenius law can be used to relate the transition rates with these energies like
\begin{equation}
\label{eq:physgamma}
\begin{aligned}
  \gamma_u &\propto \exp\left[-\beta \Delta U\right] \\
  \gamma_d &\propto \exp\left[-\beta (\Delta U+ a F)\right] \,,
\end{aligned}
\end{equation} 
where $\beta=(k_B T)^{-1}$ is the inverse temperature. From Eq.~\eqref{eq:coeff} follows
\begin{equation}
\label{eq:drift}
\begin{aligned}
  V&= a(\gamma_u+\gamma_d)\frac{\gamma_u-\gamma_d}{\gamma_u+\gamma_d}\\
  &=a(\gamma_u+\gamma_d)\tanh(\beta a F/2)\\
  &\simeq \frac{a^2}{2}\beta(\gamma_u+\gamma_d)F\,,
\end{aligned}
\end{equation}
the last approximation holding if $aF$ is small with respect to the typical thermal fluctuations. In this case, we have an explicit expression for the mobility
\begin{equation}
  \mu=\frac{V}{F} \simeq \frac{a^2}{2}\beta(\gamma_u+\gamma_d)\,.
\end{equation} 
Comparing this result to Eq.~\eqref{eq:coeff} one gets
\begin{equation}
 \mu=\beta D\,,
\end{equation} 
i.e. Einstein's relation, which is indeed expected to hold at equilibrium.

The characterization of the coefficients $\mu$ and $D$ is the goal of the measurement strategies proposed in the following sections.

\subsubsection{Stationary profile}
\label{sec:statprof}
The stationary solutions to Eq.~\eqref{eq:evrho} are of the form
\begin{equation}
\label{eq:stat1}
    \rho(z)=\frac{1}{2a} + \frac{D}{\sigma Va}\tanh \cbr{\frac{z-z_0}{\sigma}}\,,
\end{equation}
where $z_0$ and $\sigma$  are free parameters that are fixed by  the boundary conditions of the problem: $z_0$ represents the center of the transition front between the bright and dark region of the ion cloud and depends on the ratio of bright and dark ions; $\sigma$ is a typical width of this transition front.

 By substituting Eq.~\eqref{eq:stat1} in the density current given by Eq.~\ref{eq:current}  one finds that in the stationary state, the density current is
$$
J(z)=\frac{V}{4a}-\frac{D^2}{aV\sigma^2}\,.
$$
Because the stationary state is an equilibrium state where no density currents are present, we  impose the additional condition $J=0$, leading to 
\begin{equation}
\sigma=\frac{2D}{V}    
\end{equation}
 hence
\begin{equation}
\label{eq:statprof}
    \rho(z)=\frac{1}{2a}\sbr{1 + \tanh \cbr{\frac{V}{2D}(z-z_0)}}\,.
\end{equation}

The constant $z_0$ is fixed by  the  normalization on the ion density. Denoting by $N_b$ the number of bright ions, the relation
\begin{equation}
\label{eq:normcond}
    \int_{0}^Ldz \rho(z)=N_b
\end{equation}
leads to
$$
\frac{L}{2a}+\frac{D}{aV}\ln \sbr{\cosh\cbr{\frac{LV}{2D}}-\tanh\cbr{\frac{Vz_0}{2D}}\sinh\cbr{\frac{LV}{2D}}}=N_b\,,
$$
hence, $z_0$ depends on  $\sigma=2D/V$ as
\begin{equation}
z_0=\sigma\tanh^{-1}\sbr{\frac{\cosh\cbr{L/\sigma}-e^{\cbr{2aN_b-L}/\sigma}}{\sinh\cbr{L/\sigma}}}\,.
\end{equation}
If the front width $|\sigma|$ is much smaller than the total length $L$, we can approximate the argument of the $\tanh^{-1}$ as
$$
\frac{\cosh\cbr{L/\sigma}-e^{\cbr{2aN_b-L}/\sigma}}{\sinh\cbr{L/\sigma}} \simeq \begin{cases}
  1-2e^{2(aN_b-L)/\sigma }  \quad &\text{if }\sigma>0\\
   2e^{2aN_b/\sigma }-1  \quad &\text{if }\sigma<0\,.
\end{cases}
$$
By recalling the small-$x$ expansion$$
\tanh^{-1}(1-x)\simeq -\frac{1}{2}\log\cbr{\frac{x}{2}}-\frac{x}{4}+O(x^2)\,,
$$
$z_0$ takes the simple form:
\begin{equation}
\label{eq:z0_app1}
    z_0\simeq \begin{cases}
        L\cbr{1- \frac{N_b}{N}} + o\cbr{\sigma} \quad &\text{if }V>0\\
        \frac{N_b L}{N} + o\cbr{\sigma} \quad &\text{if }V<0\,.
    \end{cases}
\end{equation}

\subsubsection{Ellipsoidal geometry}
\label{sec:ellipse}

The experimental setup described in Section~\ref{sec:exp} confines the ion cloud in an ellipsoidal shape. It is experimentally verified that the action of the forcing laser does not alter this geometry. It is thus natural to model the dynamics as taking place in a 3d lattice with reflecting boundaries on an ellipsoidal domain. Along the $z$ axis, the dynamics is the one described in the previous paragraphs. Along the $xy$ plane, since no external forces are exerted, we expect self-diffusion.

The evolution equation for the density of ions is now given by
\begin{equation}
\label{eq:evrho3d}
\partial_t \tilde{\rho} = - V\partial_z\tilde{\rho} +2a^3V\tilde{\rho} \partial_z \tilde{\rho} + D(\partial^2_x+\partial^2_y+\partial^2_z)\tilde{\rho}\,,
\end{equation} 
where $\tilde{\rho}$ is a volume density, justifying the different dimensional factor in front of the nonlinear term, with respect to Eq.~\eqref{eq:evrho}. Since, for every fixed value of $z$, the dynamics on the accessible domain of the $xy$ plane is purely diffusive, the stationary solution is a density profile $\tilde{\rho}(x,y,z)$, whose dependence on the $x$ and $y$ variables is only due to the constraint of the confining potential:
\begin{equation}
\label{eq:statprof3d}
    \tilde{\rho}(x,y,z)=\Theta\sbr{4  R^2\cbr{\frac{z}{L}-\frac{z^2}{L^2}}-x^2-y^2}\frac{\rho(z)}{a^2}\,,
\end{equation}
where $\Theta(\cdot)$ is the Heavyside step-function, and  $R$ the transversal semi-axis of the ellipsoid. The linear density $\rho(z)$ is defined by Eq.~\eqref{eq:stat1}, where the parameter $z_0$ needs to be fixed by taking into account the new geometry of the system.  By recalling that the area of the section perpendicular to the $z$ axis measures
\begin{equation}
\label{eq:envelope}
    S(z)=4 \pi R^2\cbr{\frac{z}{L}-\frac{z^2}{L^2}}\,,
\end{equation}
the density of bright ions projected to the $z$ axis is 
\begin{equation}
\label{eq:ellprofile}
\begin{aligned}
\tilde{\rho}_{meas}(z)&=\int dx\,dy\,\tilde{\rho}(x,y,z)\\
&=S(z) \rho(z)\\
&=\frac{2 \pi R^2}{a^3}\cbr{\frac{z}{L}-\frac{z^2}{L^2}}\sbr{1 + \tanh \cbr{\frac{V}{2D}(z-z_0)}}\,.
\end{aligned}
\end{equation}
This is the quantity that is actually measured in experiments.
The value of $z_0$ can be fixed again by imposing the normalization condition~\eqref{eq:normcond}. One gets the relation
\begin{equation}
\label{eq:z0_app2}
    3\int_0^L  \frac{dz}{L}\cbr{\frac{z}{L}-\frac{z^2}{L^2}}\sbr{1+\tanh \cbr{\frac{V}{2D}(z-z_0)}}=\frac{N_b}{N}\,,
\end{equation}
which can be inverted numerically to find the value of $z_0$, given $V/D$ and the density of bright ions (see Fig.~\ref{fig:z0} in the next Section).

\subsection{Dynamics with a finite lifetime of the dark state}
\label{sec:shelving}
So far we have neglected the shelving dynamics and have assumed that the bright or dark state of each ion is fixed during the observational time. In this section we  relax this hypothesis, assuming that the ions can switch from bright to dark, and vice-versa, during the dynamics. When variations of this state occur much more frequently than the typical displacements, a uniform distribution of bright ions along the $z$ axis is expected to be observed. If, instead, the characteristic dark state lifetimes  are of the same order of the ones responsible for the swapping of neighbour ions, nontrivial competing effects are expected to arise. 

Let us consider again the simple one-dimensional model of Eq.~\eqref{eq:evw} where we now include the possibility that a bright ion becomes dark with rate $\Gamma_d$, and a dark ion becomes bright with rate $\Gamma_b$. The shelving process is induced  by  laser excitation and thanks to a coherent three photon process, these two rates can be tuned independently \cite{champenois06,collombon19b}. The discrete lattice model becomes then
\begin{equation}
\begin{aligned}
    \frac{d p_n}{dt}
    &=\Gamma_b(1-p_n)-\Gamma_d p_n\\
    &-\gamma_u p_n (1-p_{n+1})-\gamma_d p_n (1-p_{n-1})\\
 &+\gamma_u p_{n-1} (1-p_{n})+\gamma_d p_{n+1} (1-p_{n})\,,
 \end{aligned}
\end{equation}
leading to the continuous-space evolution equation
\begin{equation}
\label{eq:evrhoshelv}
\partial_t \rho = \frac{\Gamma_b}{a} -(\Gamma_b+\Gamma_d)\rho- V\partial_z\rho +2aV\rho \partial_z \rho + D\partial^2_z\rho\,.
\end{equation}

The stationary state for the density cannot be expressed in closed form.  However,  some quantitative predictions can still be made concerning this state. First of all, the average number of bright ions is controlled by the relation 
\begin{equation}
N_b=\frac{\Gamma_b}{\Gamma_d + \Gamma_b}N
\end{equation}
which results from a balance between bright and dark ions in the stationary state.

For a force exerted by the laser on the bright ions oriented toward the positive direction of the $z$ axis and for a transition region size which is smaller than the whole cloud, the density is expected to be close to $a^{-1}$ (only bright ions) when $z$ approaches $L$ and to $0$ (only dark ions) when $z \simeq 0$. In the former regime, the stationary state  coming from Eq.~\eqref{eq:evrhoshelv} can be approximated  as
\begin{equation}
\label{eq:bound1}
    \partial_z \rho(L) = \frac{\Gamma_d}{aV}\,;
\end{equation}
and in the latter case by
\begin{equation}
\label{eq:bound2}
    \partial_z \rho(0) = \frac{\Gamma_b}{aV}\,.
\end{equation}

We thus get a normalization condition for the density profile and an approximation for its behavior close to the cloud boundaries. This information is useful when devising diffusivity measurement strategies in the non-negligible shelving regime. Other useful information are also obtained by numerical simulations of the system as ruled by Eq.~\eqref{eq:evrhoshelv} as reported in Appendix~\ref{sec:shelvingsim}. A rich phenomenology of bright ion density profiles can be observed  when varying two scaling parameters: the ratio $\theta=(\gamma_u-\gamma_d)/(\gamma_u+\gamma_d)$ which scales the unbalance between the swapping probabilities and $\varepsilon=(\Gamma_b+\Gamma_d)/(\gamma_u+\gamma_d)$ which scales the  probability of state exchange relative to the mean swapping probability.

\section{Experimental validation}\label{sec:compare_exp}

\begin{figure}
    \centering
    \includegraphics[width=\linewidth]{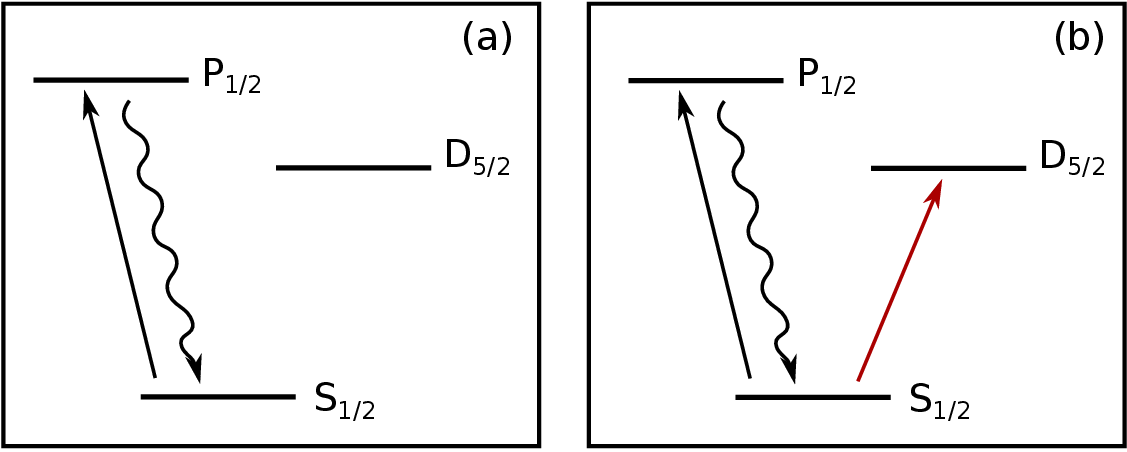}
    \caption{Scheme of the electronic levels of Ca$^+$ involved in the two regimes considered in the experiment. Panel (a) represents the cycle of excitation (black straight arrow) and spontaneous emission (black wavy arrow) in the laser-cooling process. Panel (b) refers to the regime in which shelving is present  also (excitation to the shelved state is represented by the red arrow).}
    \label{fig:levels}
\end{figure}

In this section, we present an experimental validation  of the model in the frozen limit ($\varepsilon \ll 1$) described in section.~\ref{sec:frozen}. In practice, to go from the situation of Fig.~\ref{fig:picture}a to Fig.~\ref{fig:picture}b, we leave the Doppler cooling lasers, and an additional repumping laser that depopulates the D$_{3/2}$ state, on and we add an extra laser beam at 729~nm that drives part of the ions into the D$_{5/2}$ in a metastable state (shelving process)~\cite{nagourney86}, like sketched  in Fig.~\ref{fig:levels}. The lifetime of the metastable state D$_{5/2}$ is 1.2~s \cite{guan15}, which is far longer than the swapping times $1/\gamma_u$ and $1/\gamma_d$.
The experimental data are extracted from the pictures of the laser induced fluorescence emitted by a cloud of ions, with and without shelving like explained in section~\ref{sec:exp}. The data processing allows to extract the characteristics of the ellipsoid formed by the picture of the cloud and to deduce the cloud sizes $L$ and $R$ based on the pixel dimension, which is 13~$\mu$m and the optical magnification measured to  $12.6\pm 0.1$. By defining the boundary of cloud picture, it is possible to integrate the signal in the other direction transverse to the force direction and compute an integrated signal $I(z)$. By taking care of setting the experimental conditions to have a linear response of the photon counting pixels and a uniform laser beam intensity over the whole cloud, the integrated signal $I(z)$ can be used as a proxy for the density of bright ions $\rho(z)$, with a scaling factor $\eta$, taking into account the probability for each ion to scatter photons and the detection efficiency and detector gain. The signal integrated over the whole ellipsoid is proportional to the number of bright ions in the cloud. The ratio of the two signals collected with and without shelving of part of the ions in a dark state gives access to $N_b/(N_b+N_d)$. In the case of Fig.~\ref{fig:picture}, this ratio is $0.58 \pm 0.02$, resulting in a ratio $\Gamma_b/\Gamma_d=1.38\pm0.04$.

\begin{figure}
    \centering
    \includegraphics[width=.9\linewidth]{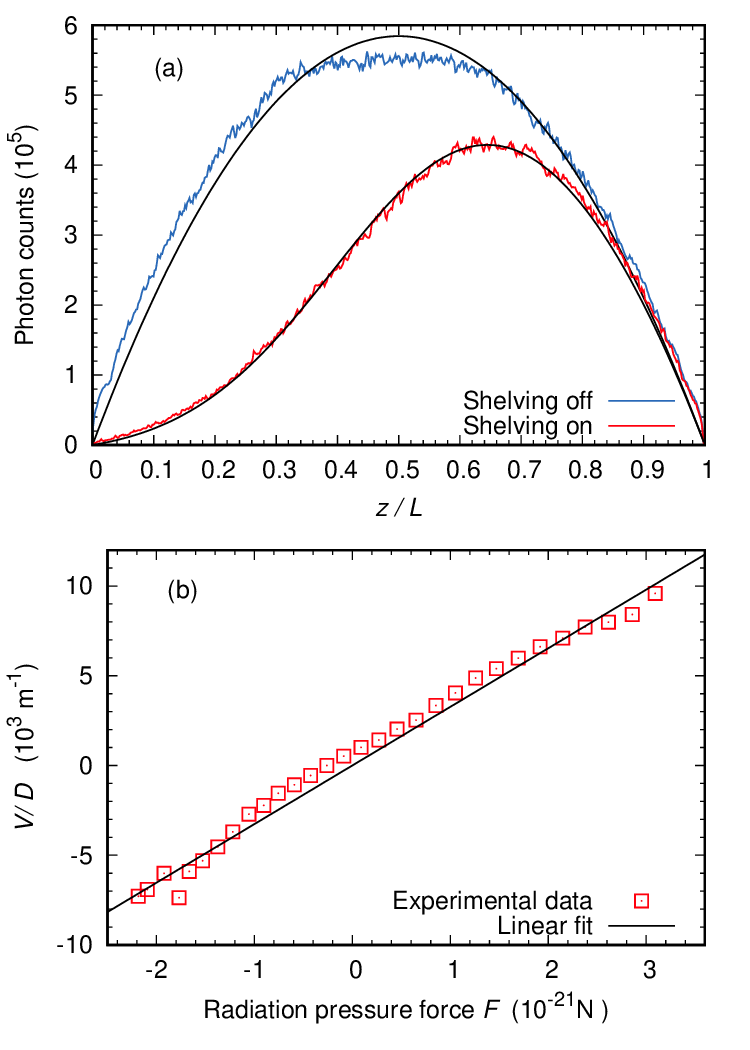}
    \caption{Panel (a) : data analysis of a picture like shown in Fig~\ref{fig:picture}, for a cloud made of 1240 $\pm$ 60 ions with total length  $L=576 \pm 14 ~\mu$m, and a measurement time of 1~s. Blue curve : number of collected photons  $I(z)$, as a function of the relative  coordinate $z/L$ along the cloud symmetry axis, in the absence of shelving, when all particles are ``bright''. The black line is the best fit with a parabolic profile, which is expected because the projected density of ions is just proportional to the transversal section of the cloud, Eq.~\eqref{eq:envelope}. Red curve : number of collected photons  $I(z)$, as a function of the relative  coordinate $z/L$, in the case where part of the ions are shelved in a metastable state, so that they do not scatter photons.  The black curve is a fit which obeys Eq.~\eqref{eq:ellprofile}, taking into account the scaling factor given by the previous fit. Panel (b) shows the value of $V/D$,  obtained from this fit, for several signal profiles $I(z)$ obtained with different values of the effective pressure force $F$.  }
    \label{fig:fit_profiles}
\end{figure}

In Fig.~\ref{fig:fit_profiles}a, the integrated signal $I(z)$ is plotted as a function of $z/L$, in the absence or presence of shelving of part of the ions in a dark state, for the same cloud. In both cases, $I(z)$ is expected to obey  Eq.~\ref{eq:ellprofile} scaled with $\eta$, with $V=0$ in the absence of shelving. In this case, a 1-parameter fit of the parabolic profile allows to defines  the common prefactor $\eta 4\pi R^2/(2a^3)$. Therefore, in the case of shelving, the fit of the functional form Eq~\eqref{eq:ellprofile} only includes two free parameters, namely $z_0$ and $D/V$. In Fig.~\ref{fig:fit_profiles}b the inverse of the latter is plotted for different values of the radiation pressure force exerted by the cooling laser. To keep the temperature of the sample constant along the experiment, the laser cooling excitation process is set in the linear regime and the cooling laser beam is split in two counter-propagating beams and the balance between the two beam intensity is tuned to reach a variable effective pressure force with a conserved number of scattered photons.  The method used to  estimate the value of the effective force is explained in Appendix~\ref{sec:appforce}.  The linear behaviour observed in Fig~\ref{fig:fit_profiles}b is in agreement with the theoretical prediction, assuming a constant temperature. Indeed,
from Eqs.~\eqref{eq:coeff} and the linear approximation of Eq.~\eqref{eq:drift}, one easily obtains
\begin{equation}
    \frac{V}{D}=\frac{F}{ k_B T}\,.
\end{equation}
From  a linear fit of the plot in Fig.~\ref{fig:fit_profiles}b it is then possible to evaluate  the temperature of the system, which  in this case is
$$
T \simeq 22 \ {\rm mK}\,.
$$
Given the Wigner-Seitz radius estimated from the density to 13.8~$\mu$m, this value gives a plasma parameter $\mathcal{G}_p=55$, which defines a liquid phase. This temperature is consistent with the linear approximation of Eq.~\eqref{eq:drift} as  $\beta aF/2\simeq 0.05$ for a force of $2\times 10^{-21}$ N.  Furthermore, it is compatible with the typical temperatures reported for ion clouds where only a fraction of the ions are laser cooled \cite{schiller03,ostendorf06}.  The same method applied to other sets of measures (not shown here), with different values of the density, shows the same linear dependence and leads to values of the temperature of 19~mK and 16~mK. For a precise and accurate estimation of this temperature, all the cause of uncertainty  concerning the detection efficiency and the force estimation must be evaluated and reduced.

\begin{figure}
    \centering
    \includegraphics[width=0.9\linewidth]{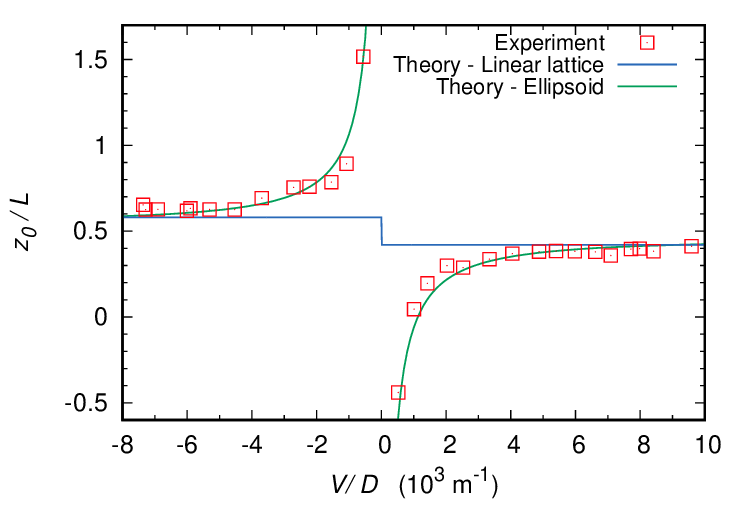}
    \caption{Transition front coordinate $z_0/L$ as a function of the drift. Red squares represent the values measured in experiments (i.e., from the fits of density profiles, as shown in Fig.~\ref{fig:fit_profiles}a. The blue solid line shows the approximate estimate provided by Eq.~\eqref{eq:z0_app1} for the linear-lattice geometry. The green solid line is computed taking into account the ellispoidal geometry, see Eq.~\eqref{eq:z0_app2}. Fraction of bright ions: $N_b/(N_b+N_d)=0.58 \pm 0.02$.}
    \label{fig:z0}
\end{figure}

As an additional check of the validity of our model, we plot in Fig.~\ref{fig:z0} the values of $z_0$ obtained from the fits of the density profiles against $V/D$, and we compare them with our theoretical prediction~\eqref{eq:z0_app2}. The agreement between the two is a further confirmation of the relevance of our description, even for situations where a front is not visible on the picture, like for the smallest values of $|V/D|$. This comparison shows also that the simplified relation~\eqref{eq:z0_app1}, which does not take into account the ellipsoidal geometry of the system, provides a fair approximation for large enough values of the external force.

The proposed model, based on a lattice-like description of the system, turns out to provide meaningful predictions in the considered cases, even if the ion cloud is in a liquid regime where a crystalline structure is not present. This is not completely surprising. The main ingredient for the exclusion-process model is the presence of a strong \textit{local} order, and this is guaranteed as soon as $\mathcal{G}_p$ exceeds a few tens~\cite{samaj2011}: in this case the pair correlation function $g(r)$ is known to exhibit a well-defined first peak, from which a first layer of neighbors can be identified~\cite{damasceno2012}. When deriving the model no assumption was made, instead, on the long-range order of the system, which is of course missing in the liquid phase.

\section{Strategies to the direct measurement of ion diffusivity}
\label{sec:strategies}
The analysis proposed in the previous section shows that it is not possible to extract definitive information about the diffusivity of the particles by only looking at static measurements  in the regime where the probability for the particle to change state is negligible. While the ratio $k_B T$ between $D$ and the mobility $\mu$ can be inferred  and compared with the expected value of the local temperature, no conclusion can be drawn about $D$ alone.

In order to circumvent this issue, hereafter we propose two possible strategies that may be pursued in future experiments. The first one still considers the frozen regime, but it focuses on dynamical measures of the density profile, acquired with fast frequency. The second one does not involve dynamical measurements, but it requires that the shelving rates are of the same order as the displacement rates.

Before proceeding with the discussion of methods to measure the diffusivity, we compare here the proposed model to the already available experimental results.
\subsection{Dynamical measurements at high frequency}

Let us assume that the dark state has an infinite lifetime (frozen regime) and that the system is prepared in the stationary state described by Eq.~\eqref{eq:statprof}, with a radiation pressure force pointing toward positive $z$. The density profile of the bright ions is characterised by $\bar{z}$, the mean value of their position defined as:
\begin{equation}
    \bar{z}=\frac{1}{N_b}\int_{0}^L dz \, \rho(z) z\,.
\end{equation}
At time $t=0$ the two counter-propagating cooling lasers are tuned in such a way that $\gamma_u=\gamma_d$, i.e. $V=0$, so that the following evolution is purely diffusive.
The dynamics of $\bar{z}$ is obtained from Eq.~\eqref{eq:evrho} and reads
\begin{equation}
\begin{aligned}
     \partial_t \bar{z} &= D\int_0^L dz \, \partial_z^2\rho(z) z\\
     &=\frac{DL}{N_b}\partial_z \rho(L)-\frac{D}{N_b}\sbr{\rho(L)-\rho(0)}\,.
\end{aligned}   
\end{equation}

If the initial condition is  characterized by a small front width compared to the total length $L$, the bright ion density profile is flat at the boundaries, and verifies
$$
\rho(0) \simeq 0 \quad \rho(L) \simeq \frac{1}{a}\,,
$$
 This means that at the beginning of the relaxation evolution, until the shape of the profile changes significantly, one has
\begin{equation}
    \partial_t \bar{z} \simeq - \frac{D}{N_b a}\,.
\end{equation}
The diffusivity can be thus measured from the variation of $\bar{z}$ by repeating the same protocol with an increasing waiting time.
\subsection{Static profile for finite lifetime of the dark state}

As discussed in section~\ref{sec:shelving}, provided that the system can be brought to a regime where the rates of bright to dark  are comparable with those of the ion displacements, the slope of the density profile close to the boundaries (Eqs.~\eqref{eq:bound1} and~\eqref{eq:bound2}) give access to $V$, provided $\Gamma_d$ and $\Gamma_b$ have been measured previously. In the corresponding case where $\Gamma_d$ and $\Gamma_b$ can be neglected (frozen regime), the fit of the density profile gives access to the front length $\sigma=2D/V$ from which one can deduce $D$. The control on $\Gamma_d$ and $\Gamma_b$ can be reached by tuning the atom-laser interaction parameters in the three-photon process detailed in \cite{champenois06}.

It would be tempting to estimate also the ratio $D/V$ from the fit in the regime where the lifetime of the dark state is finite, skipping the preliminary measurement of $\sigma$ in the frozen one. However, one must take into account that the width of the transition region is given, in this case, by the non-trivial combination of two effects: (i) the self-diffusion of the ions, which we aim at measuring; (ii) the dynamical effect due to the fact that one ion can be suddenly shelved (or unshelved), and start travelling from one side of the system to the other one. When the shelving rates are high enough, the latter effect results in a broadening of the transition region between the two coexisting phases (dark and bright) of the cloud, which does not coincide with $2D/V$ anymore. 

\begin{figure}
    \centering
    \includegraphics[width=\linewidth]{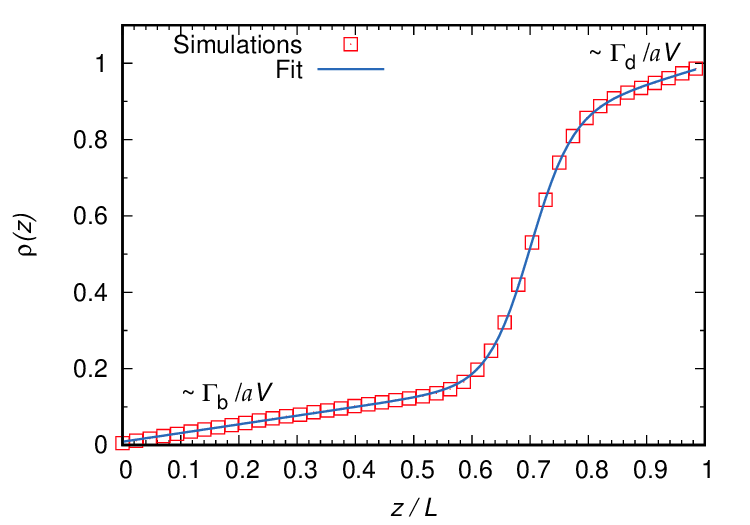}
    \caption{Numerical simulation of a case with non-negligible shelving. The stationary profile is reported for a 1-dimensional lattice model composed of $L=128$ sites. The blue solid curve is obtained as a fit on the phenomenological law~\eqref{eq:pheno}. Parameters:    
    $\Gamma_d=2\cdot 10^{-3} s^{-1}$, $\Gamma_b= 10^{-3} s^{-1}$, $\gamma_u=1.25 s^{-1}$, $\gamma_d=0.75 s^{-1}$. Average over $5\cdot 10^5$ measures of the system's histogram, at $1 s$ time-intervals. Time-step for the evolution: $\Delta t =10^{-2} s$.}
    \label{fig:shelving}
\end{figure}

At a practical level, the measurement would require two steps.
First, one should perform the experiment in a regime where the internal state of the ions is fixed. By repeating the analysis discussed in Sec.~\ref{sec:statprof}, it is possible to measure the value of $\sigma$, which relates the diffusivity $D$ to the drift $V$. Then the shelving laser parameters should be switched to reach a regime where $\Gamma_b$ and $\Gamma_d$ are not negligible, to measure them by fitting the stationary profile of the bright ions distribution.
It can be useful to fit it with the phenomenological law:\begin{equation}
\label{eq:pheno}
    \rho(z) \simeq (a_1+b_1 z) + (a_2+b_2 z)\tanh\cbr{\frac{z-z_0}{c}}\,,
\end{equation}
where it is understood that the profile has been already normalized by the local volume of the ellipsoid, in order to make it comparable with the 1-dimensional case. Although Eq.~\eqref{eq:pheno} is not an exact solution of the stationary state for the considered model, it is expected to catch the essential qualitative features of the profile when the shelving effect is small. In particular it reproduces the expected linear behaviour far from the transition region, i.e. when the value of the hyperbolic tangent is almost constant (and equal to $-1$ or $1$ depending on which boundary is considered). The frozen limit is recovered when $b_1$ and $b_2$ are equal to zero. An example from numerical simulations is shown in Fig.~\eqref{fig:shelving}.

The values of $b_1$ and $b_2$ obtained by the fit are nothing but the slopes appearing in the l.h.s. of Eqs.~\eqref{eq:bound1} and~\eqref{eq:bound2}. Since $\Gamma_d$ and $\Gamma_b$ can be measured independently by a spectroscopic method, the two equations provide an estimate for $V$. The ratio $\sigma$ having been previously measured, the diffusivity follows as
$$
D=\sigma V/2\,.
$$

\section{Conclusion}
\label{sec:concl}
In this paper, we have shown how different aspects of the atom-laser interaction can be used to measure the self-diffusion coefficient of ions within a trapped laser-cooled ion cloud. The measurements rely on 
a model developed to describe the external dynamics of the ions, when a spatial segregation is induced by the radiation pressure that is encountered only by ions that are not trapped in a dark state. The validation of this model allows to measure the temperature of the sample and to propose several protocols that should give access to other relevant parameters, like the self-diffusivity coefficient (not directly measured in the present work). These protocols rely on the control of the internal state dynamics that is permitted by a coherent multi-photon process. Together with the control gained on the sample temperature by Doppler laser cooling, these atom-laser interaction processes should allow to measure the self-diffusivity coefficient for strongly coupled non neutral-plasma, for a broad range of plasma parameters.

\begin{acknowledgments}
The authors acknowledge the contribution of Marie Houssin in the experimental preparation of the ion cloud and its interaction with the involved lasers. This work was supported by the LABEX Cluster of Excellence FIRST-TF (ANR-10-LABX-4801), within the Program ``Investissements d’Avenir'' operated by the French National Research Agency (ANR), and by the ANR through grant ANR-18-CE30-0013. MB was supported by ERC Advanced Grant RG.BIO (Contract No. 785932).

\end{acknowledgments}

\appendix

\section{Estimation of the laser's pressure force}
\label{sec:appforce}
The radiation pressure force encountered by the ions is due to the recoil associated with the absorption of photons from the laser beam $\hbar \vect{k}_L$, where $\vect{k}_L$ is the wave vector of the laser ($ k_L=2 \pi/\lambda_L$, $\lambda_L=397$ nm), pointing to the direction of the beam. In the limit of a low saturation of the excited transition, the stimulated emission following this absorption can be neglected and the only emission process is spontaneous. The averaged recoil over thousands of emission is then null and we only take into account the absorption induced recoil. The mean force depends on the number of absorption/emission cycles per unit time $\Gamma_{e} P_{e}$, with $P_{e}$ the probability for an ion to be in the excited state,  and $\Gamma_{e}$ the probability for an ion in the excited state to decay to the ground state. This spontaneous decay rate  is the inverse of the excited state lifetime, which is $\tau_e=6.9 \cdot 10^{-9}$~s for Ca$^+$ ions \cite{hettrich15}.

For a single laser beam, the mean pressure force exerted by the laser is thus given by~\cite{metcalfbook}: 
\begin{equation}
    \vect{F}=\hbar \vect{k}_L \Gamma_{e} P_{e}\,.
\end{equation}
For the measurement of $V/D$ like shown on Fig.~\ref{fig:fit_profiles},  the effective pressure force needs to be tuned. Tuning $P_e$ would induce a modification of the laser cooling efficiency and thus of the temperature. To keep the temperature constant over the experimental run, the laser beam is split in two counter-propagating beams $+\vect{k}_L$ and $-\vect{k}_L$ with a shared intensity $x_+I_L$ and $x_- I_L$ with $I_L$ the total laser intensity, and $x_{\pm}$ the tuning parameters. Again in the limit of low transition saturation, we can assume a linear response of the ions to the laser excitation and consider that $P_e=P_{e_+}+P_{e_-}$ with $P_{e_{\pm}}$ scaling with each laser beam intensity $x_{\pm}$. The average force on the $z$ axis is thus equal to
\begin{equation}
     F = \hbar k_L  (x_+-x_- ) \Gamma_{e} P_{e}\,,
\end{equation}
The probability $P_e$ can be estimated from the number $N_{e}$ of photons emitted in a given time interval $\tau_{meas}$ by the whole cloud, and the total number of ions $N$:
\begin{equation}
    P_e=\frac{\tau_e N_e}{\tau_{meas} N}\,.
\end{equation}
This quantity is basically the ratio between the time spent by a single ion in an excited state and the measurement time, and it represents therefore the probability searched for. Estimating $N_e$ requires to take into account the efficiency of the photon detector and this is certainly the largest source of uncertainty in the described protocol. In the considered experimental setup, based on an intensified CCD camera, an efficiency of 1  detected photons out of 10 has be evaluated. We obtain an estimate of about $2\cdot10^9$ emitted photons per second (slightly fluctuating) in the measurements without shelving, with a cloud of 1240 ions, resulting in $P_e \simeq 0.011$.

\section{Effect of the shelving on the stationary profile}
\label{sec:shelvingsim}

\begin{figure*}[ht]
    \centering
    \includegraphics[width=\linewidth]{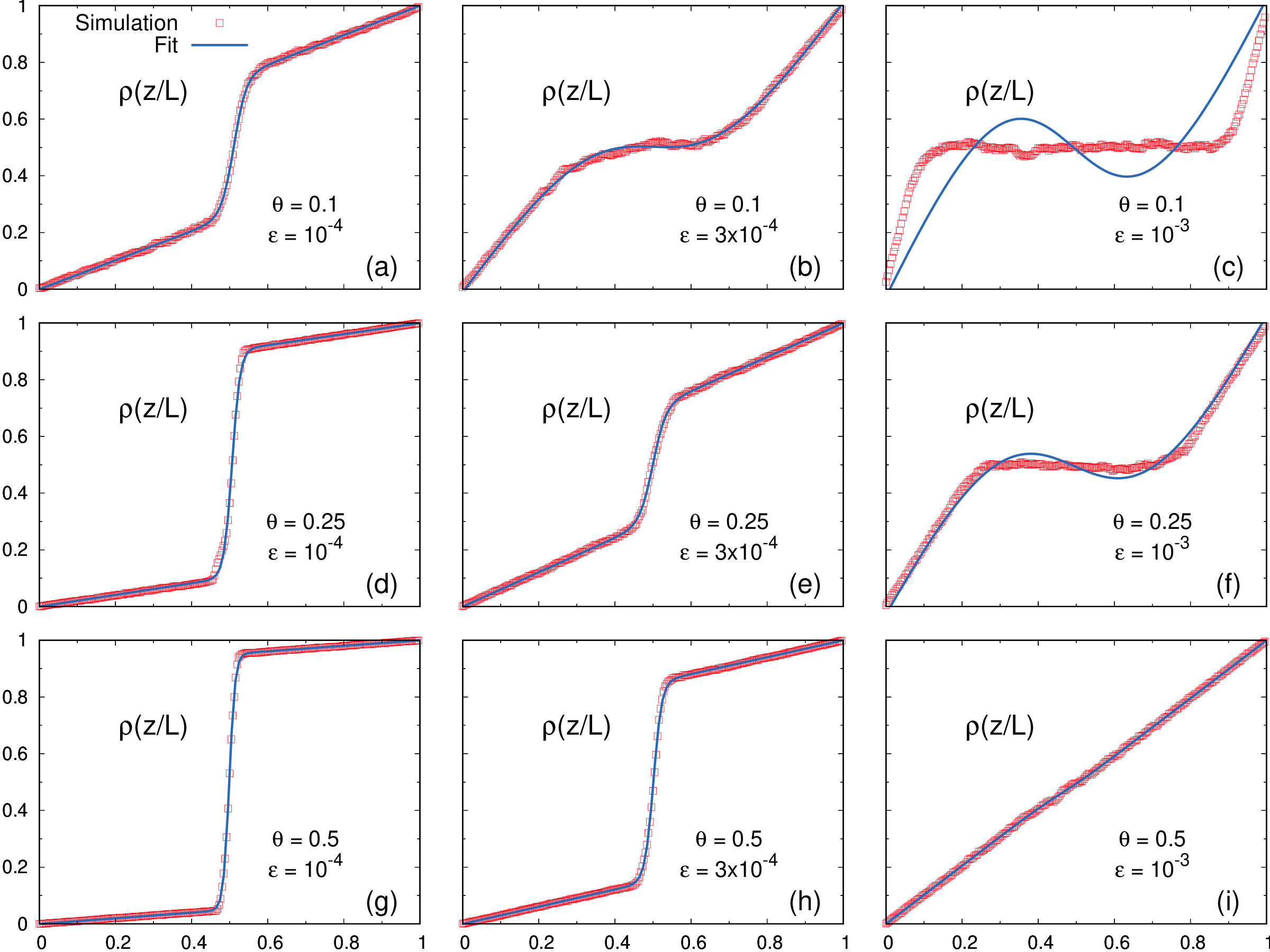}
    \caption{Numerical simulation of the stationary density profile of bright particles $\rho(z/L)$, in the case of non-negligible shelving rate, for different values of the dynamical parameters. Here a one-dimensional lattice is considered, with $N=10^3$ sites. The ratios $\theta=(\gamma_u-\gamma_d)/(\gamma_u+\gamma_d)$ and $\varepsilon=(\Gamma_b+\Gamma_d)/(\gamma_u+\gamma_d)$ are varied in the different panels (the former increases when going from top to bottom, the latter from left to right). Red squares represent the empirical profile from the numerical simulations, blue curves the best fit obtained with the functional form in Eq.~\eqref{eq:pheno}. We chose the parameters of the simulations in such a way that $\gamma_u+\gamma_d=2 s^{-1}$ and $\Gamma_b=\Gamma_d$. Other parameters as in Fig.~\ref{fig:fit_profiles}.}
    \label{fig:profiles}
\end{figure*}

The stationary density profile of a system described by Eq.~\eqref{eq:evrhoshelv} shows a quite rich phenomenology depending on the values of the dynamical parameters. Such a stationary solution does not admit a closed analytical form, but it can be explored by mean of numerical simulations. In Fig.~\ref{fig:profiles} some examples are shown for varying values of the parameters of the experiment. If the shelving rate is small enough compared to the displacement rates, the functional form provided by Eq.~\eqref{eq:pheno} approximates pretty well the measured curve. We recall that the functional form~\eqref{eq:pheno} is purely phenomenological and it is not expected to fit the profile for any choice of the parameters. Fig.~\ref{fig:profiles} shows indeed that the agreement gets worse when the ratio between the shelving and the displacement rates increases.

\bibliography{biblio}

\end{document}